\documentclass[prl,showpacs,aps,twocolumn]{revtex4}%
\usepackage{amsfonts}%
\usepackage{amsmath}%
\usepackage{amssymb}%
\usepackage{graphicx}

\begin{document}
\preprint{ }

\bigskip
\title{Unexpected features of quantum degeneracies in a pairing model with two integrable limits}

\author
{J. Dukelsky$^{1}$, J. Oko{\l}owicz$^{2}$, and M. P{\l}oszajczak$^{3}$}

\affiliation{
$^{1}$ Instituto de Estructura de la Materia, CSIC, Serrano 123, 28006 Madrid, Spain\\
$^{2}$ Institute of Nuclear Physics, Radzikowskiego 152, PL-31342 Krak\'ow, Poland\\
$^{3}$ Grand Acc\'{e}l\'{e}rateur National d'Ions Lourds (GANIL), CEA/DSM -- CNRS/IN2P3, BP 5027, F-14076 Caen Cedex 05, France \\
}

\date{\today}

\begin{abstract}
The evolution pattern of level crossings and exceptional points is
studied in a non-integrable pairing model with two different
integrable limits. One of the integrable limits has two
independent parameter-dependent integrals of motion. We
demonstrate, and illustrate in our model, that quantum
integrability of a system with more than one parameter-dependent
integral of motion is always signaled by level crossings of a
complex-extended Hamiltonian. We also find that integrability
implies a reduced number of exceptional points. Both properties
could uniquely characterize quantum integrability in small Hilbert spaces.
\end{abstract}


\bigskip

\maketitle

The search for fingerprints of the chaotic/regular dynamics in the quantum regime
is often focused on studies of spectral properties of the quantum systems. In this
context, spectral fluctuations were intensely studied for various quantum systems.
These studies lead to the BGS conjecture \cite{Boh84} that in the semiclassical limit the
spectral fluctuations of chaotic systems are described by random matrix theory.
For quantum integrable systems, Berry and Tabor \cite{Ber77} showed that the spectral
fluctuations are well described by a Poisson statistic. While chaotic systems are
characterized by level repulsion between successive levels, levels of integrable systems
are uncorrelated allowing crossings between states of the same symmetries.

Level crossings and degeneracies are important for the understanding of spectral
fluctuations \cite{Ber81} and the onset of quantum chaos \cite{Ber85}. Much effort
has been devoted to studies of degeneracies associated with
avoided crossings in quantal spectra, focusing mainly on the topological structure of the
Hilbert space and the geometric phases \cite{Ber84,Lau94}. Among these degeneracies,
one finds a diabolic point where two Riemann sheets of eigenvalues touch each
other \cite{Ber84,Ber84a}, and an exceptional point (EP) \cite{Kat95,Zir83,Hei91} where
the two sheets are entangled by the square-root type of singularity.
EP appears in the complex $g$ plane of a generic Hamiltonian $H(g)=H_0+g H_1$, 
where both $H_0$ and $H_1$ are hermitian and $[H_0,H_1] \neq 0$.
In many-body systems, EPs have been studied in schematic models like the Lipkin 
model \cite{Hei91a} and the interacting boson model along the line connecting 
the dynamical symmetries U(5) to O(6) \cite{Cej06}. Both models
can be considered as particular examples of 2-level boson pairing models 
pertaining to the general class of integrable Richardson-Gaudin (RG) models 
\cite{Duk04, Ort05}. In spite of the fact that level crossings of
eigenstates with the same global symmetries are only allowed for integrable 
systems, 2-level pairing models have no level crossing and all degeneracies 
take place in the complex plane as EPs.

In this work, we will introduce a prototypical quantum integrable system, 
the 3-level RG model, to discuss the appearance of level crossings and 
EPs, and their evolution both with complex parameters of the Hamiltonian and
with a non-integrable complex perturbation. On this basis, we prove 
that integrable models with at least two parameter-dependent integrals of motion 
(IMs) have a level crossing in the complex-extended parameter space,
providing a clear signal of their integrability. Furthermore, the inclusion 
of a non-integrable perturbation splits each level crossing into two EPs 
transforming dramatically the topology of the Hilbert space close to the
level crossing.

Let us begin by briefly reviewing the RG models which are based on the 
$SU(2)$ algebra with elements $K^+_l$, $K^-_l$, and $K^0_l$, fulfilling the 
commutation relations: $[K^+_l,K^-_{l^{\prime}}]=\delta_{ll^{\prime}} K^0_l~$
, $[K^0_l,K^{\pm}_{l^{\prime}}]=\pm\delta_{ll^{\prime}} K^{\pm}_l~$. The indices 
$l,l^{\prime}$ refer to a particular copy from a set of $L$, $SU(2)$ algebras. 
Each $SU(2)$ algebra possesses one quantum degree of freedom.
Therefore, a quantum integrable model  requires the existence of $L$ 
independent, global operators that commute with each other. These operators, 
which need not be hermitian, are the IMs. In the following, we will work with
the rational family of RG models whose IMs are \cite{Duk04}:
\begin{align}
R_{l}=K_{l}^{0}& +4g\sum_{l^{\prime}\left( \neq l\right) }\frac{1}{%
\varepsilon _{l}-\varepsilon _{l^{\prime}}}\left[ \frac{1}{2}\left(
K_{l}^{+}K_{l^{\prime}}^{-}+K_{l}^{-}K_{l^{\prime}}^{+}\right) \right.
\notag \\
& ~~~~~~~~+\left. K_{l}^{0}K_{l^{\prime}}^{0}\right], \label{Rop}
\end{align}
where $g$ and $\varepsilon_l$ are $L+1$ arbiratry parameters.  
The IMs (\ref{Rop}) satisfy $[R_i,R_j]=0$ for all pairs $i,j$.

There is a profound relation between quantum integrability in finite systems 
and the existence of level crossings. In the context of the 6-sites Hubbard 
model, this problem has been addressed by Yuzbashyan {\it et al.} \cite{Yuz} 
who showed that a level crossing implies the existence of two independent 
parameter-dependent IMs. Here we complete this analysis by showing that a 
system with at least two parameter-dependent IMs has of necessity a level 
crossings in the complex plane.

Let us assume a Hamiltonian $H(g)$ depending linearly on a parameter $g$. $H(g)$ itself
is the parameter-dependent IM. If $g$ is complex then the Hamiltonian and eventually other
parameter-dependent IMs will be non-hermitian. The parameter-dependent IM $Q(g)$
commuting with the Hamiltonian: $[H(g),Q(g)]=0$, will be independent of the
Hamiltonian if it cannot be expressed as an entire function of $H(g)$ and $g$, i.e.,
$Q(g)\neq f(g, H(g))$. Let us assume that $n$ is the dimension of the Hilbert space and
$E_1(g),\cdots,E_n (g)$ and $q_1(g),\cdots,q_n (g)$ are
the corresponding eigenvalues in the basis in which both operators are diagonal.
If $Q(g)$ is an entire function of $H(g)$, then it can be expanded for any complex $g$ value as:
\begin{eqnarray}
q_1 (g)&=&a_n E^{n-1}_1(g)+\cdots +a_2 E_1 (g) + a_1   \nonumber\\
\vdots  \label{algebra}\\
 q_n (g)&=&a_n E^{n-1}_n(g)+\cdots
+a_2 E_n (g) + a_1 .\nonumber
\end{eqnarray}
This set of equations has always a solution unless for some value $g=g_0$ a pair of
equations $\{k,k^{\prime}\}$ have $E_k(g_0)=E_{k^{\prime}}(g_0)$, but
$q_k(g_0)\neq q_{k^{\prime}}(g_0)$, implying a double degeneracy in the
Hamiltonian but not in the second IM. 

The minimal rational RG model  (\ref{Rop})
allowing for level crossings should have at least three SU(2) copies. The reason is
that the sum of the IMs (\ref{Rop}) is a global conserved symmetry:
$K^0=\sum_{i=1} ^LK^0_i$, commuting with all IMs and independent of the parameter
$g$. Hence, we are left with two parameter-dependent IMs and, as shown above, at
least two parameter-dependent IMs are required for having level crossings.

In what follows, we will use the pair representation of the SU(2) algebra 
leading to pairing Hamiltonians. The elementary operators in this representation 
are the number operators $N_j=\sum_{m}a_{jm}^{\dagger}a_{jm}$ and the
pair operators $A^{\dagger}_j=\sum_{m}a_{jm}^{\dagger}a_{j\overline{m}}^{\dagger}$, 
where $j$ is the total angular momentum and $m$ is the $z$-projection. The state 
${j\overline{m}}$ is the time reversal of ${jm}$.

The relation between the operators of the pair algebra and the generators of
the SU(2) algebra is:~
$K_{j}^{0}=\frac{1}{2}N_{j}-\frac{1}{4}\Omega _{j}~$,$~~  K_{j}^{+} =\left( K_{j}^{-}\right) ^{\dagger
}=\frac{1}{2}A_{j}^{\dagger }$,  where $\Omega_j$  is the particle degeneracy
of level $j$. With this correspondence, one can introduce an integrable 3-level
pairing Hamiltonian as:
\begin{eqnarray}
H(g)=2\sum_{i}\varepsilon_{i} R_i(g) + C \equiv \sum_{i}\varepsilon_{i}N_{i}+g\sum_{ij}A_{i}^{\dagger}A_j
 ,\label{IPH}
\end{eqnarray}
where $C$ is an irrelevant constant and $\varepsilon_i$ $(i=1,2,3)$ are the
single-particle  energies. One can see  that $H(g)$ itself is a parameter-dependent
IM. As discussed above, the sum of the IMs (\ref{Rop}) yields a parameter-independent
IM, the particle number:
$N=2\sum_{l}R_{l}(g)+\frac{1}{2}\sum_{l}\Omega_{l}~.$
The second  parameter-dependent IM can be chosen  as linearly independent from
the two other IMs.
The simplest choice is $R_{1}$.
If $\varepsilon _{1}=0$, the second parameter-dependent IM
becomes:
\begin{eqnarray}
Q(g)&=&\left[ 1+g\left( \frac{\Omega _{2}}{\varepsilon _{2}}+ \frac{\Omega_{3}}{\varepsilon _{3}}\right) \right]
\frac{N_{1}}{2}+\frac{g \Omega _{1} }{\varepsilon _{2}}\frac{N_{2}}{2}+\frac{g  \Omega_{1}}{\varepsilon
_{3}}\frac{N_{3}}{2}
\nonumber \\
&-&g\left\{ \frac{1}{\varepsilon _{2}}\left[ \frac{1}{2}\left( A_{1}^{\dagger
}A_{2}+A_{2}^{\dagger }A_{1}\right) +N_{1}N_{2}\right] \right. \nonumber \\
&+&\left. \frac{1}{\varepsilon_{3}}\left[ \frac{1}{2}\left( A_{1}^{\dagger }A_{3}+A_{3}^{\dagger }A_{1}\right)
+N_{1}N_{3}\right] \right\} \label{Qg}
 . \end{eqnarray}

The position of all degeneracies in the complex-g plane are indicated by the 
roots of the coupled equations:
\begin{eqnarray}
{\rm det}\left[H\left(  g\right)  -EI\right]  = 0~;~~~~ \frac{\partial}{\partial E} {\rm det}\left[ H\left(
g\right)  -EI\right] = 0
 . \end{eqnarray}
By eliminating $E$ from these two equations, we are left with the discriminant $D(g)$,
a polynomial in $g$ of degree  $M\leq n(n-1)$. The discriminant can be written 
as \cite{Zir83}:
\begin{eqnarray}
D(g)=\prod_{m<m^{\prime}}\left[E_m(g)-E_{m^{\prime}}(g)\right]^2 ~\ ,
\end{eqnarray}
where $E_m(g)$, $E_{m^{\prime}}(g)$ denote the complex eigenvalues of $H(g)$.
The eigenvalue degeneracies $E_m(g)=E_{m^{\prime}}(g)$ at
$g=g_{\alpha}$ ($\alpha=1,\dots,M$), can be found numerically by looking for
sharp minima of $D(g)$. It turns out that for real values of $\varepsilon_j$, the roots
are real or complex conjugate pairs. One finds two kinds of solutions in quantum
integrable models: (i) single roots corresponding to EPs that are common to all 
IMs, and (ii) double- (multiple-) root solutions which indicate non-singular sharp 
crossings of two (or more) levels with two (or more) orthogonal wave functions, 
related to the existence of at least two independent IMs.
\begin{figure}[hbt]
\begin{center}
\includegraphics[width=6cm,angle=-90]{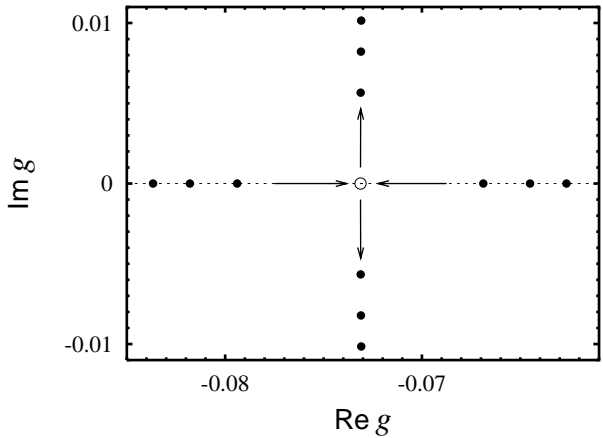}
\caption{Collision and subsequent scattering of two level crossings as a function of the
energy $\varepsilon_3$ in a complex-extended integrable 3-level pairing Hamiltonian
(\ref{IPH}). Points are plotted in a descending order of  $\varepsilon_3$. For more details, 
see the discussion in the text.}
\label{fig11}
\end{center}
\end{figure}

Fig. \ref{fig11} shows an evolution of two level crossings, the double-roots 
of $D(g)$, in the complex-$g$ plane as a function of the energy of the third 
single-particle level $\varepsilon_3$ ($\varepsilon_3>\varepsilon_2$).
The systems has 4 pairs of fermions in a valence space with level degeneracies 
$\Omega_1=6, \Omega_2=4, \Omega_3=2$ and $\varepsilon_1=0$ , $\varepsilon_2=1$. 
In the limit $\varepsilon_3 \rightarrow \infty$, this system decouples effectively into the 
two 2-level models: the first one with level $\varepsilon_3$ occupied and
the second one with $\varepsilon_3$ empty. In this limit level crossings are forbidden 
and, indeed, the two level crossings that appear for finite values of
$\varepsilon_3$ move to $\pm \infty$. With decreasing $\varepsilon_3$, two level
crossings coming from $\pm \infty$ approach each other in the real axis up to a
critical value $\varepsilon_3^{(\rm cr)}=1.8499$ where they coalesce. The level 
crossing at this point corresponds to the quadruple-root of $D(g)$. For
$\varepsilon_3<\varepsilon_3^{(\rm cr)}$ this crossing splits again into the two 
double-root level crossings which move into the
complex-$g$ plane. The presence of such level crossings in the complex plane is
a clear signature of quantum integrability, and shows the necessity of extending the
demonstration of operator independency given in (\ref{algebra}) to the whole complex
parameter space.

EPs associated to single roots of the discriminant $D(g)$, unlike level crossings, are
common to all parameter-dependent IMs including the Hamiltonian. It should be also
noted that EPs appear in the quantum integrable model even though no manifestation
of level repulsion is expected and the spectral fluctuations of the hermitian Hamiltonian
obey a Poisson distribution \cite{Rela}. In that sense, level repulsion may be a sufficient
but not a necessary condition for the appearance of EPs.

Another feature that we found in this 3-level pairing model, as
well as in more general multi-level pairing models, is the
reduction of the total number of discriminant roots whose maximum
value is $M_{max}=n(n-1)$. In the particular case shown in Fig. 1,
$M_{max}=20$ but we found 16 roots consisting of 2 level crossings
(double roots) and 12 EPs (single roots).

In the following, we generalize the 3-level pairing  Hamiltonian
(\ref{IPH}) to study effects of  non-integrability:
\begin{eqnarray}
H(g)=\sum_{i}\varepsilon_{i}N_{i}+\zeta g\sum_{ij}A_{i}^{\dagger}A_j-(1-\zeta)g\sum_{i} N_i^2   \ .
\label{NIPH}
 \end{eqnarray}
For $\zeta=1$, Eq. (\ref{NIPH}) corresponds to the integrable Hamiltonian (\ref{IPH}).
For $\zeta=0$, the Hamiltonian (\ref{NIPH}) is also integrable with the number operators
$N_i$ playing the role of parameter-independent IMs. In the interval: $0<\zeta<1$, the
Hamiltonian (\ref{NIPH}) is non-integrable, i.e. it does not possess an independent IM
other than the Hamitonian and the total number operator. Two main features characterize
the emergence of non-integrability, firstly the level crossings break into pairs of EPs 
and secondly the missing roots of the discriminant come into play as EPs from $\infty$.
\begin{figure}[hbt]
\begin{center}
\includegraphics[height=7cm,angle=-90]{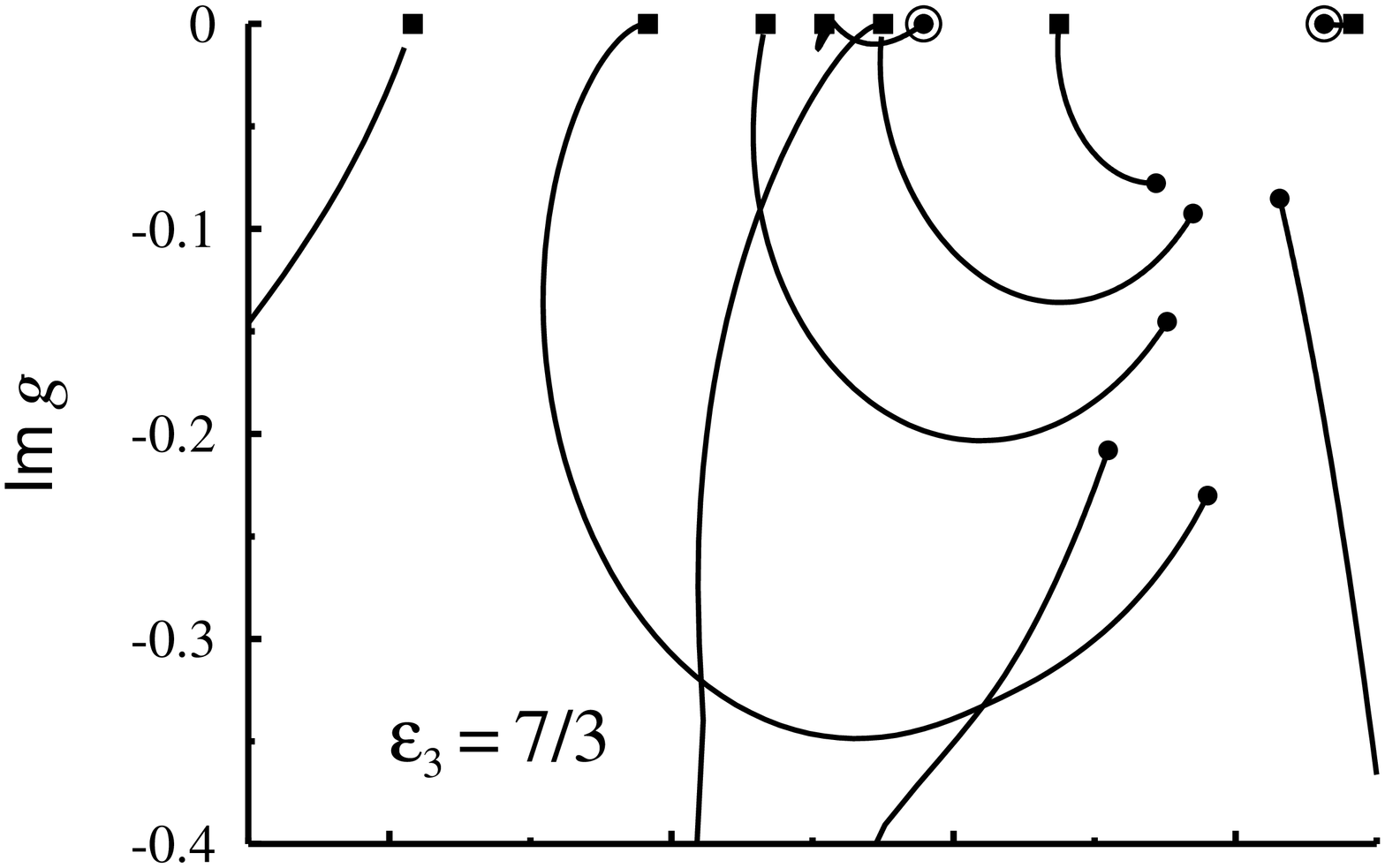}
\includegraphics[height=7cm,angle=-90]{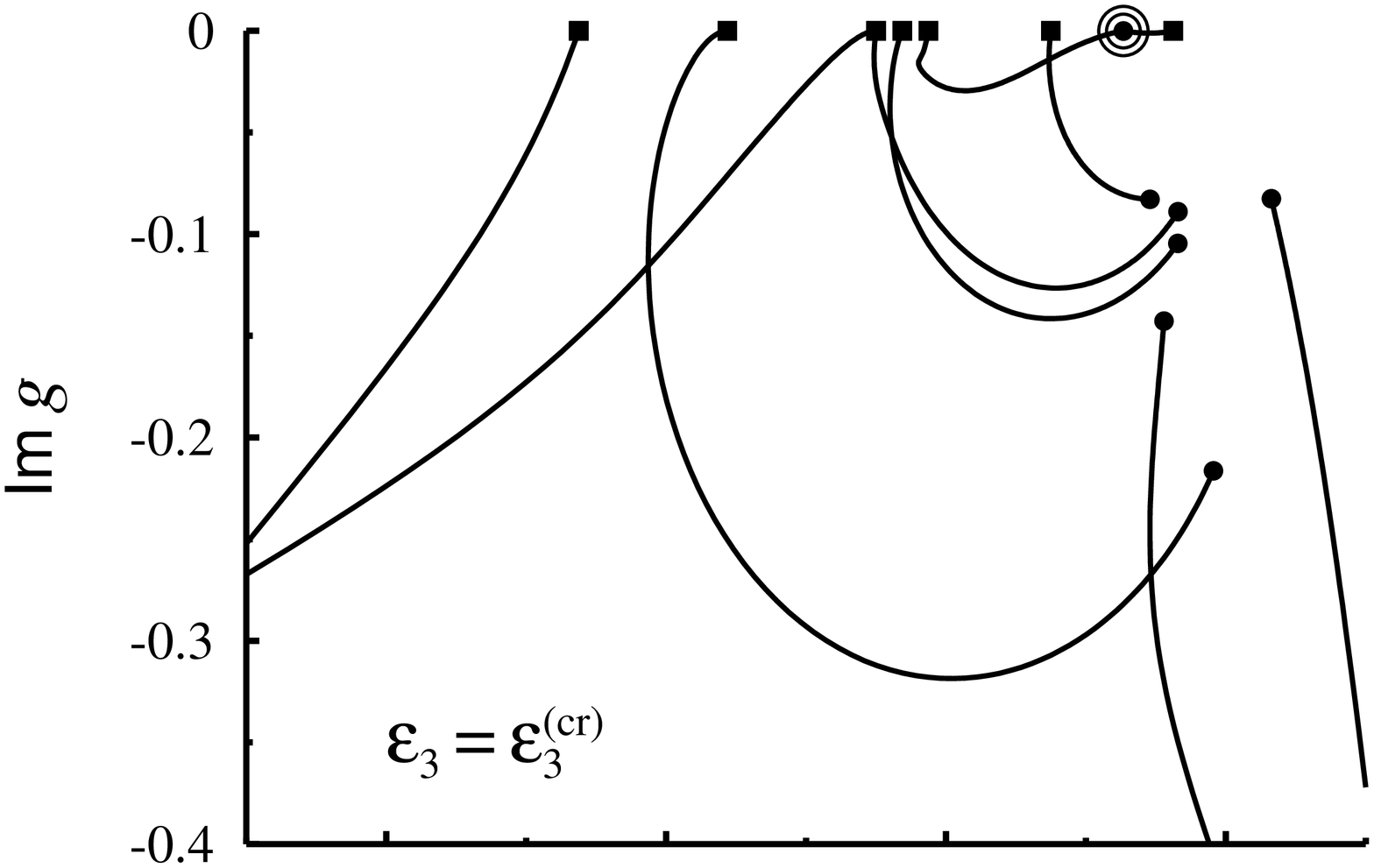}
\includegraphics[height=7cm,angle=-90]{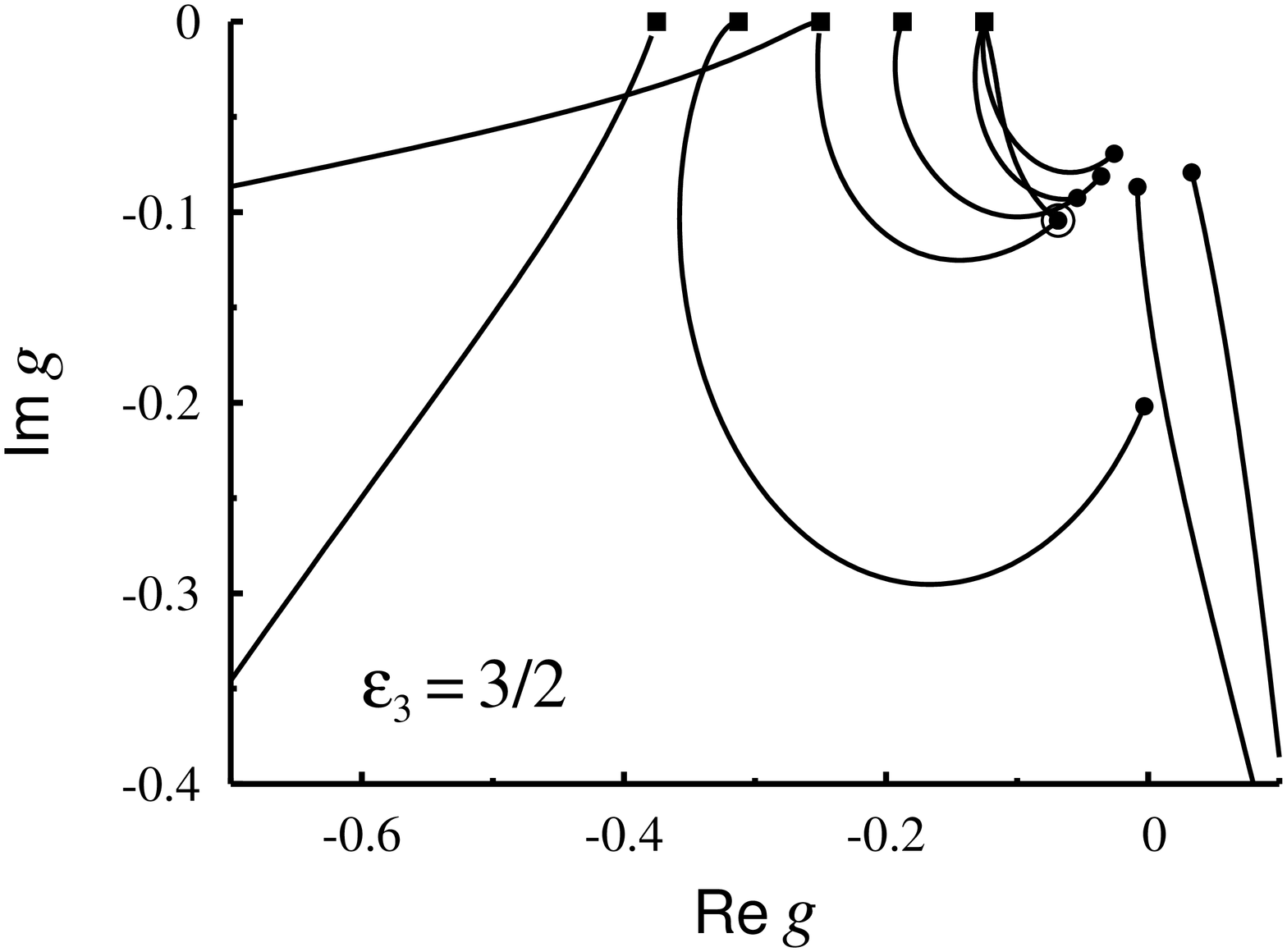}
\caption{The evolution of level crossings and EPs of the complex-extended 3-level
pairing Hamiltonian (\ref{NIPH}) in the interval $0\leq\zeta\leq1$.
Circles and squares denote the position of degeneracies at $\zeta=1$ and 0, 
respectively. The double circles depict the double-root level crossing at $\zeta=1$. 
The triple circle shows the quadruple-root level crossing corresponding to the 
coalescence of two double-root level crossings. For more details, see the caption
of Fig. \ref{fig11} and the discussion in the text.}
\label{fig25}
\end{center}
\end{figure}

Fig. \ref{fig25} shows the global pattern of level crossings and
EPs as a function of the parameter $\zeta$ ($0\leq\zeta\leq 1$)
 for the 3-level system of Fig. 1 in three regimes: $\varepsilon_3>\varepsilon_3^{(cr)}$
 ($\varepsilon_3=7/3$), $\varepsilon_3=\varepsilon_3^{(cr)}$, and
$\varepsilon_3<\varepsilon_3^{(cr)}$ ($\varepsilon_3=3/2$). Energies 
of levels '1' and '2' are fixed at: $\varepsilon_1=0, \varepsilon_2=1$.  
Only the lower half-plane of $g$ is shown where all eigenvalues are 
either discrete states on the real-$g$ axis or else decaying resonances.
Complex conjugate degeneracies situated in the upper half-plane
(${\cal I}m(g)>0$) correspond to capturing resonances. In the
limit of $\zeta=1$, the two level crossings for $\varepsilon_3=7/3$  
are located along the real-$g$ axis. One may also notice a quadruple-root 
level crossing on  the real-$g$ axis at $\varepsilon_3=\varepsilon_3^{(cr)}$ 
(see also Fig. \ref{fig11}). Moreover, one can see the location of 6 EPs of the 
integrable pairing Hamiltonian (\ref{IPH}) in all regimes of $\varepsilon_3$.

With decreasing $\zeta$, one observes several distinct effects.
Firstly, each double-root level crossing  at $\zeta=1$ breaks into
a pair of  EPs. For $\varepsilon_3=3/2$, two EPs resulting from this
fragmentation follow independent trajectories in the complex-$g$
plane and end up in two different level crossings at $\zeta=0$.
For $\varepsilon_3=7/3$, the level crossing  at $\zeta=1$ breaks into two EPs 
symmetrically with respect to the real-$g$ axis. Since this symmetry is conserved 
for all $\zeta$, these EPs end up in the same level crossing for $\zeta=0$. 
Secondly, roots that are missing in the integrable limit ($\zeta=1$) appear from 
$g=\infty$ and end up in level crossings for $\zeta \rightarrow 0$.
Thirdly, in the limit  $\zeta\rightarrow 0$,  all EPs either collapse
in different level crossings at the real-$g$ axis or escape to infinity. 
The double level crossings  found in this limit correspond to the two 
different eigenvalue degeneracies. For $\varepsilon_3=3/2$, 
one can see three EPs converging on each side of the real-$g$ axis to a 
single point. At this sextuple-root of the discriminant, one finds a sharp 
crossing of three eigenvalues.

The degree of the discriminant, which for $\zeta=0,1$ equals 16 in
all regimes of $\varepsilon_3$, becomes $M=n(n-1)=20$ in the
non-integrable regime. Notice the absence  of EPs in the
integrable case $\zeta=0$ reflecting the fact that the Hamiltonian
is diagonal in the original basis for any $g$ value. 

In conclusion, we have shown that a system with two independent
parameter-dependent IMs has at least one level crossing in the
complex parameter space. We have used a minimal integrable
pairing model consisting of three levels to exemplify this issue.
Moreover, we have found that the degree of the discriminant is
reduced in this integrable limits, giving rise to a lower number
of EPs, but still a fraction of them persists in spite of the fact
that there is no manifestation of level repulsion. If
integrability is broken by the inclusion of a non-integrable
perturbation, all level crossings split into EPs and other EPs
come into the complex plane from $\infty$ recovering the maximal
degree $M_{max}=n(n-1)$ of the discriminant. We conjecture that these
two unexpected properties, level crossings in the complex-$g$
plane and the reduction in the number of EPs, uniquely define a
quatum integrable system. If this conjecture proves to be true,
then it might be particularly useful in studies of finite systems
with small dimensional Hilbert spaces, where the analysis of
spectral fluctuations is unreliable. More work has to be done to
elucidate the relation between generic integrable Hamiltonians and
the missing roots of the discriminant in order to confirm this conjecture.

We acknowledge fruitful discussions with C. Esebbag. This work was supported in part 
by the Spanish MEC under
grant No. FIS2006-12783-C03-01 and by the CICYT(Spanish)-IN2P3(French) cooperation.

\end{document}